\documentclass[twocolumn, showpacs,amsmath,amssymb,aps,prl,floatfix, 10pt]{revtex4-1}
\usepackage[latin1]{inputenc}
\usepackage[american]{babel}
\usepackage[T1]{fontenc}
\usepackage{mathrsfs}
\usepackage{hyperref}
\usepackage{multirow}
\usepackage{bm, epsfig}
\usepackage{graphicx}
\usepackage{subeqnarray}
\usepackage{bbold}
\usepackage{upgreek}
\usepackage{tikz}
\usepackage{slashed}
\usepackage{lineno}

\usepackage{dcolumn}
\usepackage{physics}
\usepackage{nicefrac}
\newcolumntype{w}[1]{D{.}{.}{#1}}
\newcommand{\Za}{Z\alpha}
\newcommand{\vare}{\varepsilon}
\newcommand{\balpha}{\vec{\alpha}}

\hypersetup{colorlinks=true, urlcolor=blue, citecolor=blue, filecolor=blue, linkcolor=blue}

\newcommand{\ThreeJ}[6]{
  \left(
    \begin{array}{ccc}
      #1  & #2  & #3 \\
      #4  & #5  & #6 \\
    \end{array}
  \right) }
\newcommand{\half}{\frac12}

\usepackage{changes}
\definechangesauthor[name={Natalia}, color=magenta]{nat}
\definechangesauthor[name={Vladimir}, color=orange]{vay}

\usetikzlibrary{decorations.pathmorphing}
\usetikzlibrary{shapes}
\usetikzlibrary{shapes.geometric}

\tikzset{
    magnetic/.style={
        fill,
        shape border rotate=0,
        isosceles triangle,
        isosceles triangle apex angle=60,
        node distance=1,
        minimum height=.1
    }
}

\tikzset{
    othermagnetic/.style={
        fill,
        shape border rotate=90,
        isosceles triangle,
        isosceles triangle apex angle=60,
        node distance=1,
        minimum height=.1
    }
}

\begin{document}

\title{QED calculations of the nuclear recoil effect in muonic atoms}

\date{\today}

\author{Vladimir~A.~Yerokhin}
\affiliation{Max-Planck-Institut f\"{u}r Kernphysik, Saupfercheckweg 1, 69117 Heidelberg, Germany}
\author{Natalia~S.~Oreshkina}
\email[Email: ]{Natalia.Oreshkina@mpi-hd.mpg.de} 
\affiliation{Max-Planck-Institut f\"{u}r Kernphysik, Saupfercheckweg 1, 69117 Heidelberg, Germany}

\begin{abstract}
The nuclear recoil effect, known also as the mass shift, is one of 
theoretical contributions to the energy levels in muonic atoms.
Accurate theoretical predictions  are therefore needed for extracting e.g. the nuclear charge radii from experimental spectra. 
We report rigorous QED calculations of the nuclear recoil correction in muonic atoms, carried out to all orders in the nuclear binding strength parameter $Z\alpha$ (where $Z$ is the nuclear charge number and $\alpha$ is the fine structure constant).
The calculations show differences with the previous approximate treatment of this effect, most pronounced for the lowest-lying bound states. The calculated recoil correction was found to be sensitive to the nuclear charge radius, which needs to be accounted for when extracting nuclear parameters from the measured spectra. 
\end{abstract}

\maketitle


\section{Introduction}

Muonic atoms are a class of atomic systems where a negatively charged muon replaces an electron in the atomic cloud. 
Unlike electrons, muons penetrate about 200 times deeper into the atomic nucleus due to their larger mass.
This phenomenon allows one to use muonic atoms as a powerful tool to investigate the inner structure 
and properties of atomic nuclei \cite{Wheeler1949, Feinberg:1963xj, borie:82}.
In combination with theoretical predictions of spectral properties of muonic atoms, experimental
studies of the transition energies provide accurate determinations of the nuclear charge radii, 
the magnetic dipole and the electric quadrupole moments of the nuclei 
\cite{Pohl2010, Piller:1990zza, SCHALLER1980333, DEY1979418, RUETSCHI1984461}. New generation of
these experiments is currently being implemented in the Paul Scherrer Institute \cite{muX, re185q2020},
which requires complementary advances on the theory side.

The previous-generation theory did not always succeeded in adequately describing the experimental data. 
In particular, there are unexplained disagreements observed in muonic Pb, Zr, and Sn atoms \cite{Phan:1985em, Piller:1990zza,Yamazaki:1979tf,Bergem:1988zz}. 
At the time, these disagreements were ascribed to the insufficiently known nuclear polarization (NP) corrections, which were typically treated as free fitting parameters  when interpreting the experimental spectra. However, recent studies of the NP correction \cite{valuev:22,haga:02} did not support this supposition and called for systematic {\em ab initio} QED recalculations of spectra of muonic atoms \cite{Oreshkina_muse} and a search for alternative solutions \cite{beyer2023challenging}. 

In the present work we report the first rigorous QED calculation of the nuclear recoil effect in muonic
atoms. Previously, the nuclear recoil calculations in muonic atoms were 
performed approximately \cite{michel:17}, with a partial inclusion
of relativistic effects. 
This is in contrast to the case of electronic atoms, where the nuclear recoil was studied
rigorously withing QED. In particular,
formulas valid to all order in the electron-nucleus coupling strength parameter
$\Za$ were derived by Shabaev~\cite{shabaev:85,shabaev:98:rectheo}
and calculated numerically in Refs.~\cite{artemyev:95:pra,artemyev:95:jpb}. 
These formulas obtained for the
electronic atoms can in principle be applied also to muonic atoms, but
the point-nucleus model assumed in the derivation is not an adequate
approximation for muonic atoms. One therefore needed a generalization of the nuclear
recoil treatment for the extended nuclear size.

A partial inclusion of the finite nuclear size (fns) into the nuclear recoil was reported
in Refs.~\cite{shabaev:98:jpb,yerokhin:15:recprl}. The complete treatment, however,
required a derivation of the generalized photon propagator, which
was accomplished only recently in Ref.~\cite{pachucki:23:prl}. 
In the present work we use the general expressions for the fns nuclear recoil effect
derived in these studies to perform rigorous numerical 
calculations for muonic atoms without any expansion in the parameter $\Za$. 

Relativistic units ($\hbar = c = 1$) and the Heaviside charge units $\alpha = e^2/(4\pi)$
are used throughout the paper. 

\section{Theory}

General formulas for the nuclear recoil correction to energies of hydrogen-like atoms
were derived by Shabaev~\cite{shabaev:85,shabaev:98:rectheo}.
These formulas are
valid to all orders in the nuclear binding strength parameter
$\Za$. They were derived for the point nuclear model;
within this model they
can be applied both for the electronic and the muonic atoms. 
The expression for the nuclear recoil correction 
to the energy of a bound lepton in a state $a$ is
\begin{align}\label{1}
E_{\rm rec} =&\, \frac{i}{2\pi M}\, \int_{-\infty}^{\infty}d\omega\,
\sum_n
 \frac1{\vare_a + \omega - \vare_n(1-i0)}
\nonumber \\ & \times
 \bra{a} \vec{p} - \vec{D}(\omega) \ket{n}
   \bra{n} \vec{p} - \vec{D}(\omega) \ket{a}\,,
\end{align}
where $M$ is the nuclear mass,
$\vec{p}$ is the momentum operator, $\vare_a$ is the Dirac energy of the state $a$,
$\vec{D}(\omega)$ is
obtained from the transverse part of the photon propagator in the Coulomb gauge $D_C^{ij}$ by
$$
D^j(\omega) = -4\pi Z\alpha \, \alpha^i \, D_C^{ij}(\omega,\vec{r})\,
$$
and $\alpha^i$ are the Dirac matrices. The summation over $n$ in Eq.~(\ref{1}) is carried out over the
complete Dirac spectrum. The transverse part of the photon propagator in the Coulomb gauge
can be expressed in terms of the scalar function ${\cal D}(\omega,r)$ as
\begin{align}
D_{C}^{ij}(\omega,\vec{r}) = \delta^{ij}\,{\cal D}(\omega,r) + \frac{\nabla^i\nabla^j}{\omega^2} \, \Big[ {\cal D}(\omega,r)
 - {\cal D}(0,r)\Big]\,. 
\end{align}
In case of the standard photon propagator describing the interaction between the two point-like particles, the function 
${\cal D}(\omega,r)$ has a simple form \cite{shabaev:98:rectheo}
\begin{align}\label{3}
{\cal D}(\omega,r) =
 -\frac{e^{i|\omega|r}}{4\pi r}\, .
\end{align}

It was recently demonstrated \cite{pachucki:23:prl} that the formula (\ref{1}) 
can be generalized to describe the nuclear recoil effect
for an extended-size nucleus if one replaces the standard photon propagator by
the generalized photon propagator describing the interaction between
a point-like and an extended-size particle. 
The derivation of the generalized photon propagator presented in Ref.~\cite{pachucki:23:prl} 
yields the following result for the function ${\cal D}(\omega,r)$:
\begin{align}\label{4}
{\cal D}(\omega,r) = \int \frac{d^3k}{(2\pi)^3}\, e^{i\vec{k}\cdot\vec{r}}\,\frac{\rho({\vec k}^2-\omega^2)}{\omega^2-{\vec k}^2}\,,
\end{align}
where $\rho(q^2)$ is the charge formfactor of the nucleus in momentum space. 
An interesting feature of the
generalized photon propagator is that it requires knowledge of the charge formfactor $\rho(q^2)$ not only for the
positive but also the negative arguments $q^2$. Generally, for calculating the nuclear recoil correction 
with the propagator
(\ref{4}), we need the analytical continuation of the charge formfactor 
into the whole complex plane of momenta.

Following Ref.~\cite{pachucki:23:prl}, we use the exponential model of the nuclear charge distribution. Within 
this model,
the nuclear charge density in coordinate and momentum space is given by
\begin{align}\label{5a}
\rho(r) = \frac{\lambda^3}{8\pi}\,e^{-\lambda r}\,,\ \  \
\rho({\vec k}^2) = \frac{\lambda^4}{(\lambda^2+{\vec k}^2)^2}\,,
\end{align}
respectively, where the parameter $\lambda$ is expressed in terms of 
the root-mean-square nuclear charge radius $r_C$ as 
$\lambda = 2\sqrt{3}/r_C$. Within this parametrization
of the nuclear charge distribution, the extended-size photon propagator is written in coordinate space as
\cite{pachucki:23:prl}
\begin{align}
{\cal D}(\omega,r) =
 -\frac1{4\pi}
  \bigg[
  \frac{e^{i|\omega|r}}{r} - \frac{e^{i\sqrt{\omega^2-\lambda^2}r}}{r}
    - \frac{i\lambda^2}{2} \frac{e^{i\sqrt{\omega^2-\lambda^2}r}}{\sqrt{\omega^2-\lambda^2}}
  \bigg]\,.
\end{align}

For numerical calculations,
it is convenient to separate the nuclear recoil correction (\ref{1}) 
into several parts, which are induced by the exchange of arbitrary number of Coulomb photons, by the
Coulomb and one transverse photon ($E_{\rm tr1}$), and by the
Coulomb and two transverse photons ($E_{\rm tr2}$). Furthermore, the Coulomb-photon part is
separated into the leading-order part $E_{\rm L}$ and the higher-order Coulomb-photon part $E_{\rm C}$.
We thus write
\begin{align}
E_{\rm rec} = E_{\rm L} + E_{\rm C} + E_{\rm tr1} + E_{\rm tr2}\,.
\end{align}
The leading-order contribution is 
\begin{align}
E_{\rm L}  = \frac{1}{2M}\,\bra{a} \vec{p}\,^2 \ket{a}\,.
\end{align}
We note that $E_L$ has a form of an expectation value of the nonrelativitic reduced-mass operator $\vec{p}\,^2/(2M)$
with the Dirac wave functions. The remaining Coulomb-photon contribution is given by
\begin{align}
E_{\rm C}  = -\frac{1}{M}\, \sum_{\vare_n < 0} \bra{a} \vec{p} \ket{n}\, \bra{n} \vec{p} \ket{a}\,,
\end{align}
where the summation is performed over the {\em negative-continuum} part of the Dirac spectra. 
The correction $E_\mathrm{C}$ is suppressed by
a factor of $(\Za)^3$ as compared to the leading-order contribution $E_{\rm L}$. 

The one-transverse-photon and the two-transverse-photon contributions are given by
\begin{align}
E_{\rm tr1} = &\,
     -\frac{i}{\pi M}\, \int_{-\infty}^{\infty}d\omega\,
\sum_n
 \frac1{\vare_a + \omega - \vare_n(1-i0)}\,
\nonumber \\ & \times
 \bra{a} \vec{p}\ket{n}\, 
   \bra{n}\vec{D}(\omega) \ket{a}\,, \\
E_{\rm tr2} = &\,
     \frac{i}{2\pi M}\, \int_{-\infty}^{\infty}d\omega\,
\sum_n
 \frac1{\vare_a + \omega - \vare_n(1-i0)}\,
\nonumber \\ & \times
  \bra{a} \vec{D}(\omega)\ket{n} \, 
   \bra{n} \vec{D}(\omega) \ket{a}\,. \\
\end{align}
The one-transverse-photon and the two-transverse-photon contributions are suppressed with respect to
the leading-order contribution $E_{\rm L}$ by factors $(\Za)^2$ and $(\Za)^3$, respectively.

\section{Numerical evaluation}

We now bring the general formulas for the nuclear recoil correction to the form suitable for the
numerical evaluation. 
The leading-order contribution $E_{\rm L}$ is calculated after transforming the matrix element as
follows \cite{shabaev:94:rec}
\begin{align}
\bra{a} \vec{p}\,^2\ket{a} = \bra{a} \big( \balpha\cdot \vec{p}\big)\,^2\ket{a} = 
\bra{a} \big( \vare_a - \beta m - V_{\rm nucl}\big)^2\ket{a}\,,
\end{align}
where $\beta$ is the Dirac matrix, $\vare_a$ is the Dirac energy of the state $a$,
and $V_{\rm nucl}(r)$ is the nuclear binding potential. 
Performing angular integration, we obtain
\begin{align}
E_\mathrm{L} = &\, \frac1{2M} \int_0^{\infty} dr\, r^2\, \Big\{
 \big[ (\vare_a-V_{\rm nucl})^2 + m^2 \big]\, \big[ g_a^2(r)+f_a^2(r) \big]
 \nonumber \\ &
 - 2m\,(\vare_a-V_{\rm nucl})\,
 \big[ g_a^2(r)-f_a^2(r) \big]\Big\}\,,
\end{align}
where $g_a(r)$ and $f_a(r)$ are the upper and the lower radial components of the wave function of the state $a$,
defined as in Ref.~\cite{rose:61}.

For calculating the remaining Coulomb-photon contribution $E_C$ it is convenient to apply the
identity  \cite{shabaev:94:rec}
\begin{align}
\vec{p} = \frac12\, \big\{ \balpha, h_D \big\} - \balpha\, V_{\rm nucl}\,,
\end{align}
where $\big\{ \balpha, h_D \big\} = \balpha\,h_D + h_D \balpha$, and $h_D$ is the Dirac Hamiltonian 
\begin{align}
h_D = \balpha \cdot \vec{p} + \beta m + V_{\rm nucl}(r)\,,
\end{align}
and $h_D\ket{a} = \vare_a\ket{a}$. Therefore,
\begin{align}
\bra{a} \vec{p} \ket{n} = \bra{a} \balpha\,\phi(r) \ket{n}\,,
\end{align}
where $\phi(r) = (\vare_a+\vare_n)/2-V_{C}(r)$. The angular integration is performed 
analytically, yielding
\begin{align}
E_{\rm C} = -\frac{1}{M} \sum_{\vare_n < 0} \frac{3}{2j_a+1}\, \Big[ R_C(an) \Big]^2\,.
\end{align}
Here, $j_a$ is the total angular momentum quantum number of the state $a$ and 
the radial integral $R_C$ is defined by
\begin{align}
R_C(an) = &\,\int_0^{\infty}dr\,r^2\,\phi(r)
 \Big[g_a(r)\,f_n(r)\,S_{10}(\kappa_a,-\kappa_n)
\nonumber \\ &
   - f_a(r)\,g_n(r)\,S_{10}(-\kappa_a,\kappa_n)
   \Big]\,,
\end{align}
where
$\kappa_i$ is the Dirac angular-momentum quantum number of the state $i$
and $S_{JL}(\kappa_1,\kappa_2)$ are the standard angular coefficients defined in Appendix~\ref{app}.

Calculations of the transverse-photon contribution is more complicated than that of $E_{\rm C}$ because of
the integration over the photon energy $\omega$. First of all, we make the Wick rotation of the
$\omega$ integration contour. This rotation yields pole terms originating from the intermediate states
more or equally deeply bound as the reference state. We obtain
\begin{align}
E_{\rm tr1} = &\,
     -\frac{2}{M}\,
\sum_{0 < \vare_n \leq \vare_a} a_n\,
 \bra{a} \vec{p}\ket{n}\,
   \bra{n}\vec{D}(\Delta_{an}) \ket{a}
\nonumber \\    &
     +\frac{2}{\pi M}\, \int_{0}^{\infty}d\omega\,
\sum_{n} \frac{\Delta_{an}}{\Delta_{an}^2+\omega^2}
 \bra{a} \vec{p}\ket{n}\,
   \bra{n}\vec{D}(i\omega) \ket{a}\,,
\end{align}
\begin{align}
&\, E_{\rm tr2} = 
     \frac{1}{M}\,
\sum_{0 < \vare_n \leq \vare_a} a_n\,
 \bra{a} \vec{D}(\Delta_{an})\ket{n}\,
   \bra{n}\vec{D}(\Delta_{an}) \ket{a}
\nonumber \\    &
     -\frac{1}{\pi M}\, \int_{0}^{\infty}d\omega\,
\sum_{n} \frac{\Delta_{an}}{\Delta_{an}^2+\omega^2}
 \bra{a} \vec{D}(\Delta_{an})\ket{n}\,
   \bra{n}\vec{D}(i\omega) \ket{a}\,,
\end{align}
where $a_n = 1$ for $\vare_n \neq \vare_a$ and $a_n = \nicefrac12$ for $\vare_n = \vare_a$.

The angular integration for the one-transverse-photon contribution is carried out analytically
using the standard Racah algebra. The result is
\begin{align}
\sum_{\mu_n}
 \bra{a} \vec{p}\ket{n}\,
   \bra{n}\vec{D}(\omega) \ket{a}
   =&\,
      \frac{3}{2j_a+1}\,R_C(an)\,\Big[ R_T^{(1)}(an)
\nonumber \\ &
       + \frac1{\sqrt{3}}\,
         C_1(\kappa_n\kappa_a)\,
         R_T^{(2)}(\omega,an)
         \Big]\,,
\end{align}
where $\mu_n$ is the momentum projection of the state $n$,
the angular coefficient $C_L(\kappa_1,\kappa_2)$ is defined in Appendix\ref{app},
and the radial integrals are
\begin{align}
R_T^{(1)}(\omega,an) = &\, \int_0^{\infty}dr\,r^2\,\Phi_1(\omega,r)
 \Big[g_a(r)\,f_n(r)\,S_{10}(\kappa_a,-\kappa_n)
\nonumber \\ & 
   - f_a(r)\,g_n(r)\,S_{10}(-\kappa_a,\kappa_n)
   \Big]\,,
\\
R_T^{(2)}(\omega,an) = &\, \int_0^{\infty}dr\,r^2\,\Phi_2(\omega,r)
 \Big[g_a(r)\,g_a(r)
   + f_a(r)\,f_n(r)
   \Big]\,.
\end{align}
Furthermore,
\begin{align}
\Phi_1(\omega,r) = &\, -4\pi\Za\, {\cal D}(\omega)\,,\\
\Phi_2(\omega,r) = &\, -4\pi\Za \frac{\vare_a-\vare_n}{\omega^2} \Big[{\cal D}^{\prime}(\omega) - {\cal D}^{\prime}(0)\Big]
\,,
\end{align}
and ${\cal D}^{\prime}(\omega) = d/(dr)\,{\cal D}(\omega,r)$.

Similarly, the integrand of the two-transverse-photon part is evaluated as
\begin{align}
\sum_{\mu_n}
 \bra{a} \vec{D}(\omega)\ket{n}\,
   \bra{n}\vec{D}(\omega) \ket{a}
   =&\,
      \frac{3}{2j_a+1}\,\Big[ R_T^{(1)}(an)
\nonumber \\ &
       + \frac1{\sqrt{3}}\,
         C_1(\kappa_n\kappa_a)\,
         R_T^{(2)}(\omega,an)
         \Big]^2\,.
\end{align}

%
%

\section{Results}

\begin{figure}[t]
\centerline{
\resizebox{0.5\textwidth}{!}{%
  \includegraphics{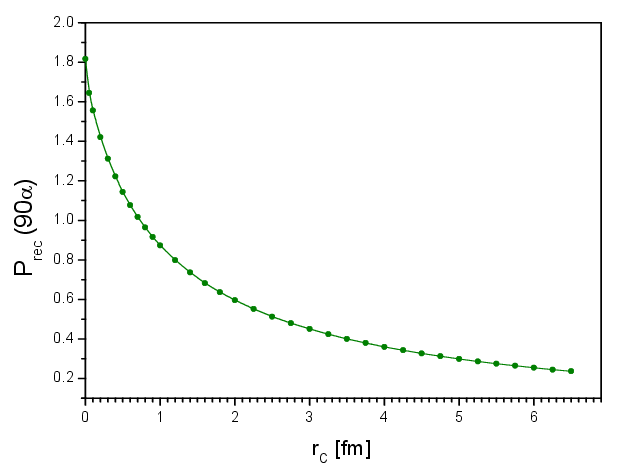}
}
}
\caption{Dependence of the nuclear recoil correction $P_{\rm rec}(\Za)$ on the
root-mean-square nuclear charge radius $r_C$, for $Z = 90$ and the $1s$ state.
\label{fig1}}
\end{figure}

We performed numerical calculations of the nuclear recoil corrections by representing the Dirac muon spectrum
with the finite basis set constructed with $B$-splines, using the dual-kinetic balance method 
\cite{shabaev:04:DKB}. The numerical procedure is very similar to that developed for the electronic
atoms in Ref.~\cite{yerokhin:16:recoil}. 
Numerical results are conveniently represented in terms of
dimensionless function $P_{\rm rec}$ defined as
\begin{align}\label{eq1}
E_{\rm rec} = m_{\mu}c^2\,\frac{m_{\mu}}{M}\,\frac{(\Za)^2}{2n^2}\,P_{\rm rec}(\Za)\,,
\end{align}
where $E_{\rm rec}$ is the recoil correction to the energy level.
In the nonrelativistic limit $\alpha \to 0$, the nuclear recoil effect reduces to the
multiplication of the binding energy by the
reduced mass prefactor. Therefore, in the nonrelativistic limit 
$P_{\rm rec}(0) = 1$.

Table~\ref{tab:rec} presents our numerical results for the nuclear recoil correction 
obtained for the $n = 1$, $n = 2$, and $n = 3$ states
of muonic atoms with the nuclear charge numbers $Z = 10$-100. For
each $Z$, we present the leading-order contribution $E_\mathrm{L}$, the complete nuclear recoil correction
$E_{\rm rec}$ calculated for the nuclear charge radius $r_C$ specified in the table, and the derivative
$E'_{\rm rec} = dE_{\rm rec}/dr_C$. 
It is interesting that the leading-order contribution $E_L$ yields a very
reasonable approximation for the complete recoil correction, even in the high-$Z$ region. E.g., 
for 
$Z = 90$ the difference between $E_L$ and $E_{\rm rec}$ does not exceed 20\%. This is in striking contrast to the electronic ions, where the corresponding difference for $Z = 90$ reaches 70\% \cite{artemyev:95:pra}. 
We conclude that for muonic atoms the QED corrections to the nuclear recoil effect are less prominent than for the electronic atoms. With the increase
of the principal quantum number $n$ and the orbital momentum $l$, we see a clear tendency that
both $E_L$ and $E_{\rm rec}$ rapidly approach the nonrelativistic limit of $P_{\rm rec}(0) = 1$. 

In the point-nucleus limit the function $P_{\rm rec}$ is the same for electronic and muonic atoms.
We checked that our numerical calculations with the point nuclear model agree with the results of 
Refs.~\cite{artemyev:95:pra,artemyev:95:jpb} for the $n = 1$ and $n = 2$ states and with those
of Ref.~\cite{anisimova:22} for the $n = 3$ states. For the extended-size nuclei, the results
for the muonic and electronic atoms are very much different. For the electronic atoms, the
finite nuclear size effect is a small correction, whereas for the muonic atoms it changes
the magnitude of the nuclear recoil effect drastically. 
Fig.~\ref{fig1} presents a plot of $P_{\rm rec}$ as a function of the nuclear radius $r_C$, for $Z=90$ 
and the $1s$ state. It is seen that the finite nuclear size effect reduces the
numerical value of the function $P_{\rm rec}$ by nearly an order of magnitude.

The dependence of $E_{\rm rec}$ on the nuclear charge radius $r_C$ for muonic atoms is 
thus quite strong. For $Z = 90$ and the $1s$ state, a 10\% change of the nuclear radius
value given in Table~\ref{tab:rec} leads to a 9\% change of $E_{\rm rec}$. In order to facilitate the
analysis of the dependence of $E_{\rm rec}$ on the nuclear radius, Table~\ref{tab:rec} presents 
results for the first derivative of $E_{\rm rec}$ on $r_c$.
The results for the derivative can be used to obtain
$E_{\rm rec}$ for nuclear radii different from the ones presented in the table. 

\begin{table*}
\caption{
Nuclear recoil correction with inclusion of the finite nuclear size for muonic atoms,
in terms of the function $R_{\rm rec}$. For each $Z$, the upper line presents results
for the leading-order recoil correction $E_{\rm L}$, whereas the second and the third lines
give the complete recoil correction $E_{\rm rec}$ and its derivative $E'_{\rm rec} = d/(dr_C)E_{\rm rec}$, in fm$^{-1}$. 
\label{tab:rec}}
\begin{ruledtabular}
\begin{tabular}{cccw{2.7}w{2.7}w{2.7}w{2.7}w{2.7}w{2.7}w{2.7}w{2.7}w{2.7}w{2.7}}
\multicolumn{1}{c}{$Z$} &
\multicolumn{1}{c}{$r_C$[fm]} & &
    \multicolumn{1}{c}{$1s$} &
    \multicolumn{1}{c}{$2s$} &
    \multicolumn{1}{c}{$2p_{1/2}$} &
    \multicolumn{1}{c}{$2p_{3/2}$} &
    \multicolumn{1}{c}{$3s$} &
    \multicolumn{1}{c}{$3p_{1/2}$} &
    \multicolumn{1}{c}{$3p_{3/2}$} &
    \multicolumn{1}{c}{$3d_{3/2}$} &
    \multicolumn{1}{c}{$3d_{5/2}$}
\\
\hline\\[-5pt]
  10 & 3.0055 & $E_\mathrm{L}$     &   0.96805 &   0.98776 &   1.00573 &   1.00174 &   0.99269 &   1.00470 &   1.00205 &   1.00160 &   1.00071 \\
     &        & $E_\mathrm{rec}$   &   0.95994 &   0.98114 &   1.00125 &   0.99994 &   0.98769 &   1.00112 &   1.00025 &   1.00029 &   1.00000 \\
     &        & $E'_\mathrm{rec}$  &  -0.02270 &  -0.01156 &  -0.00006 &  -0.00004 &  -0.00775 &  -0.00005 &  -0.00003 &  -0.00000 &  -0.00000 \\[3pt]
  20 & 3.4776 & $E_\mathrm{L}$     &   0.87756 &   0.95090 &   1.02196 &   1.00637 &   0.97030 &   1.01804 &   1.00773 &   1.00643 &   1.00285 \\
     &        & $E_\mathrm{rec}$   &   0.85493 &   0.92996 &   1.00399 &   0.99910 &   0.95400 &   1.00373 &   1.00052 &   1.00116 &   0.99997 \\
     &        & $E'_\mathrm{rec}$  &  -0.05638 &  -0.03025 &  -0.00103 &  -0.00076 &  -0.02062 &  -0.00081 &  -0.00060 &  -0.00000 &  -0.00000 \\[3pt]
  30 & 3.9283 & $E_\mathrm{L}$     &   0.75975 &   0.89972 &   1.04420 &   1.01123 &   0.93888 &   1.03650 &   1.01500 &   1.01455 &   1.00641 \\
     &        & $E_\mathrm{rec}$   &   0.72569 &   0.86303 &   1.00441 &   0.99506 &   0.90920 &   1.00489 &   0.99892 &   1.00258 &   0.99991 \\
     &        & $E'_\mathrm{rec}$  &  -0.07336 &  -0.04240 &  -0.00509 &  -0.00386 &  -0.02962 &  -0.00394 &  -0.00301 &  -0.00003 &  -0.00002 \\[3pt]
  40 & 4.2694 & $E_\mathrm{L}$     &   0.65040 &   0.84997 &   1.06504 &   1.01240 &   0.90861 &   1.05452 &   1.02086 &   1.02600 &   1.01137 \\
     &        & $E_\mathrm{rec}$   &   0.60926 &   0.79817 &   0.99720 &   0.98451 &   0.86500 &   1.00069 &   0.99292 &   1.00450 &   0.99977 \\
     &        & $E'_\mathrm{rec}$  &  -0.07614 &  -0.04787 &  -0.01363 &  -0.01051 &  -0.03440 &  -0.01030 &  -0.00809 &  -0.00017 &  -0.00012 \\[3pt]
  50 & 4.6519 & $E_\mathrm{L}$     &   0.54921 &   0.79955 &   1.07352 &   1.00411 &   0.87760 &   1.06440 &   1.02101 &   1.04057 &   1.01752 \\
     &        & $E_\mathrm{rec}$   &   0.50554 &   0.73483 &   0.97566 &   0.96297 &   0.82075 &   0.98653 &   0.97923 &   1.00663 &   0.99937 \\
     &        & $E'_\mathrm{rec}$  &  -0.07029 &  -0.04856 &  -0.02613 &  -0.02083 &  -0.03604 &  -0.01915 &  -0.01569 &  -0.00069 &  -0.00051 \\[3pt]
  60 & 4.9123 & $E_\mathrm{L}$     &   0.47194 &   0.75942 &   1.06937 &   0.98730 &   0.85370 &   1.06685 &   1.01636 &   1.05778 &   1.02451 \\
     &        & $E_\mathrm{rec}$   &   0.42713 &   0.68243 &   0.94210 &   0.93242 &   0.78339 &   0.96475 &   0.95957 &   1.00849 &   0.99841 \\
     &        & $E'_\mathrm{rec}$  &  -0.06341 &  -0.04813 &  -0.03905 &  -0.03214 &  -0.03689 &  -0.02783 &  -0.02369 &  -0.00192 &  -0.00141 \\[3pt]
  70 & 5.3108 & $E_\mathrm{L}$     &   0.39779 &   0.71411 &   1.03855 &   0.95315 &   0.82513 &   1.05278 &   1.00047 &   1.07573 &   1.03114 \\
     &        & $E_\mathrm{rec}$   &   0.35515 &   0.62843 &   0.88934 &   0.88633 &   0.74347 &   0.93068 &   0.92933 &   1.00852 &   0.99582 \\
     &        & $E'_\mathrm{rec}$  &  -0.05368 &  -0.04532 &  -0.05009 &  -0.04328 &  -0.03604 &  -0.03477 &  -0.03111 &  -0.00461 &  -0.00342 \\[3pt]
  80 & 5.4648 & $E_\mathrm{L}$     &   0.35068 &   0.68598 &   1.01045 &   0.92163 &   0.80935 &   1.04295 &   0.98798 &   1.09463 &   1.03764 \\
     &        & $E_\mathrm{rec}$   &   0.30833 &   0.58982 &   0.83977 &   0.84305 &   0.71454 &   0.89989 &   0.90175 &   1.00698 &   0.99194 \\
     &        & $E'_\mathrm{rec}$  &  -0.04800 &  -0.04448 &  -0.05784 &  -0.05159 &  -0.03650 &  -0.03972 &  -0.03665 &  -0.00835 &  -0.00621 \\[3pt]
  82 & 5.5012 & $E_\mathrm{L}$     &   0.34186 &   0.68027 &   1.00324 &   0.91434 &   0.80607 &   1.04013 &   0.98489 &   1.09824 &   1.03876 \\
     &        & $E_\mathrm{rec}$   &   0.29972 &   0.58225 &   0.82902 &   0.83366 &   0.70877 &   0.89324 &   0.89575 &   1.00628 &   0.99087 \\
     &        & $E'_\mathrm{rec}$  &  -0.04683 &  -0.04422 &  -0.05900 &  -0.05299 &  -0.03652 &  -0.04047 &  -0.03758 &  -0.00929 &  -0.00692 \\[3pt]
  83 & 5.5211 & $E_\mathrm{L}$     &   0.33747 &   0.67734 &   0.99935 &   0.91051 &   0.80436 &   1.03855 &   0.98323 &   1.10000 &   1.03929 \\
     &        & $E_\mathrm{rec}$   &   0.29546 &   0.57844 &   0.82350 &   0.82882 &   0.70584 &   0.88981 &   0.89266 &   1.00587 &   0.99028 \\
     &        & $E'_\mathrm{rec}$  &  -0.04623 &  -0.04408 &  -0.05953 &  -0.05366 &  -0.03652 &  -0.04082 &  -0.03801 &  -0.00979 &  -0.00729 \\[3pt]
  90 & 5.7848 & $E_\mathrm{L}$     &   0.30219 &   0.64980 &   0.95782 &   0.87374 &   0.78631 &   1.01813 &   0.96452 &   1.10922 &   1.04084 \\
     &        & $E_\mathrm{rec}$   &   0.26260 &   0.54731 &   0.77647 &   0.78709 &   0.68129 &   0.86019 &   0.86549 &   1.00007 &   0.98405 \\
     &        & $E'_\mathrm{rec}$  &  -0.04071 &  -0.04189 &  -0.06148 &  -0.05721 &  -0.03561 &  -0.04215 &  -0.04026 &  -0.01414 &  -0.01072 \\[3pt]
  92 & 5.8571 & $E_\mathrm{L}$     &   0.29309 &   0.64228 &   0.94545 &   0.86288 &   0.78133 &   1.01205 &   0.95897 &   1.11132 &   1.04090 \\
     &        & $E_\mathrm{rec}$   &   0.25417 &   0.53886 &   0.76302 &   0.77504 &   0.67454 &   0.85173 &   0.85765 &   0.99786 &   0.98186 \\
     &        & $E'_\mathrm{rec}$  &  -0.03929 &  -0.04131 &  -0.06172 &  -0.05792 &  -0.03537 &  -0.04236 &  -0.04072 &  -0.01550 &  -0.01181 \\[3pt]
 100 & 5.8570 & $E_\mathrm{L}$     &   0.27316 &   0.62987 &   0.92314 &   0.83949 &   0.77628 &   1.00674 &   0.95158 &   1.12470 &   1.04425 \\
     &        & $E_\mathrm{rec}$   &   0.23394 &   0.51805 &   0.72744 &   0.74374 &   0.65831 &   0.83026 &   0.83809 &   0.99208 &   0.97551 \\
     &        & $E'_\mathrm{rec}$  &  -0.03701 &  -0.04144 &  -0.06399 &  -0.06117 &  -0.03622 &  -0.04438 &  -0.04322 &  -0.02015 &  -0.01536 \\[3pt]
\end{tabular}
\end{ruledtabular}
\end{table*}

Table~\ref{tab:comparison} presents a comparison of the nuclear recoil correction
evaluated within the leading-order approximation, $E_{\rm L}$, the complete
nuclear recoil correction evaluated within QED, $E_{\rm rec}$, and the
results obtained previously within an approximate relativistic treatment
\cite{michel:17}. We observe that the previous approach yields typically
a slightly better approximation than the leading-order approximation $E_{\rm L}$.
However, the approximate results are systematically lower than the full-QED values
and the previous uncertainties underestimate the missing relativistic and 
QED effects. 

\begin{table}[t]
    \caption{Nuclear recoil correction to the energies of muonic atoms,
    in keV. $E_L$ is the leading-order contribution, $E_{\rm rec}$ is the
    complete nuclear recoil correction.}
    \label{tab:comparison}
    \centering
    \begin{tabular}{c c c w{3.5} w{3.5} w{1.5}}
    \hline
    Atom &
    $r_C$ [fm] &
    State &
    \multicolumn{1}{c}{$E_\mathrm{L}$} &
    \multicolumn{1}{c}{$E_\mathrm{rec}$} &
    \multicolumn{1}{c}{Previous \cite{michel:17}} \\
    \hline
    $^{89}_{40}$Zr & 4.2706
    & $1s_{1/2}$  &  3.735   &  3.499  &  3.21(15) \\
    && $2s_{1/2}$ &  1.220   &  1.146  &  1.09(2) \\
    && $2p_{1/2}$ &  1.529   &  1.432  &  1.42(1) \\
    && $2p_{3/2}$ &  1.454   &  1.414  &  1.40(1) \\
    && $3s_{1/2}$ &  0.580   &  0.552  &  0.53(1) \\
    && $3p_{1/2}$ &  0.673   &  0.639  &  0.64 \\
    && $3p_{3/2}$ &  0.652   &  0.634  &  0.63 \\
    && $3d_{3/2}$ &  0.655   &  0.641  &  0.64 \\
    && $3d_{5/2}$ &  0.645   &  0.638  &  0.63 \\
    \hline
    \hline
    $^{147}_{62}$Sm & 4.9892
    & $1s_{1/2}$  &  3.810   &  3.438   &   2.88(8) \\
    && $2s_{1/2}$ &  1.566   &  1.402   &   1.26(5) \\
    && $2p_{1/2}$ &  2.224   &  1.947   &   1.92(5) \\
    && $2p_{3/2}$ &  2.050   &  1.930   &   1.92(4) \\
    && $3s_{1/2}$ &  0.787   &  0.719   &   0.66(2) \\
    && $3p_{1/2}$ &  0.988   &  0.890   &   0.88(1)\\
    && $3p_{3/2}$ &  0.941   &  0.885   &   0.88(1) \\
    && $3d_{3/2}$ &  0.985   &  0.936   &   0.92(1)\\
    && $3d_{5/2}$ &  0.952   &  0.926   &  0.91(1) \\
    \hline
    \hline
    $^{205}_{83}$Bi & 5.5008
    & $1s_{1/2}$  &   3.633   &  3.179   &  2.41(6) \\
    && $2s_{1/2}$ &   1.820   &  1.554   &  1.33(4) \\
    && $2p_{1/2}$ &   2.685   &  2.212   &  2.12(3) \\
    && $2p_{3/2}$ &   2.445   &  2.226   &  2.26(1) \\
    && $3s_{1/2}$ &   0.960   &  0.842   &  0.75(3) \\
    && $3p_{1/2}$ &   1.239   &  1.062   &  1.02(3)\\
    && $3p_{3/2}$ &   1.173   &  1.065   &  1.03(3) \\
    && $3d_{3/2}$ &   1.311   &  1.199   &  1.19(2)\\
    && $3d_{5/2}$ &   1.239   &  1.180   &  1.17(2) \\
    \hline
    \end{tabular}
\end{table}

\section{Conclusion}

We have calculated the
nuclear recoil effect for muonic atoms rigorously within
QED and to all orders in the nuclear binding strength parameter $\Za$. This calculation
significantly improves the accuracy of the nuclear recoil correction as compared
to the previous treatments with partial inclusion of relativistic effects. The
results of the full QED treatment differ from the nonrelativistic
predictions for high values of $Z$ and the deeply bound states, but
rapidly converge to the nonrelativistic limit when $Z$ decreases and (or) the
highly excited states are considered. 
The nuclear recoil correction was shown to depend strongly on the nuclear charge radius.
This dependence should be taken into account when nuclear parameters are extracted
from experimental transition energies.

\appendix

\section*{Appendix: Angular coefficients $S_{JL}$ and $C_L$}
\label{app}
The coefficients $C_J(\kappa_b ,\kappa_a )$ are given by 
\begin{align} \label{CJ}
C_J(\kappa_b ,\kappa_a ) =&\, (-1)^{j_b+1/2} \sqrt{(2j_a+1)(2j_b+1)}
\nonumber \\ & \times
 \ThreeJ{j_a}{J}{j_b}{\half}{0}{-\half}
 \Pi (l_a,l_b,J) \, ,
\end{align}
where the symbol $\Pi (l_a,l_b,J)$ is unity if $l_a+l_b+J$ is even and zero
otherwise. Furthermore, $\kappa_i$ is the Dirac angular-momentum quantum number, 
$j_i = |\kappa_i|-1/2$, and $l_i = |\kappa_i+1/2|-1/2$. 

The angular coefficients $S_{JL}(\kappa_a,\kappa_b)$  are nonvanishing 
only for $L = J$, $J\pm 1$ and can be written for $J \ne 0$ as
follows:
\begin{eqnarray} \label{SJL}
S_{J\, J+1}(\kappa_a ,\kappa_b ) &=& \sqrt{\frac{J+1}{2J+1}}
 \left( 1+ \frac{\kappa_a +\kappa_b}{J+1} \right) C_J(-\kappa_b ,\kappa_a ) \ , \\
S_{J\, J}(\kappa_a ,\kappa_b ) &=& \frac{\kappa_a -\kappa_b}{\sqrt{J(J+1)}}\,
 C_J(\kappa_b ,\kappa_a )\  , \\
        \label{SJL_}
S_{J\, J-1}(\kappa_a ,\kappa_b ) &=& \sqrt{\frac{J}{2J+1}}
 \left( -1+ \frac{\kappa_a +\kappa_b}{J} \right) C_J(-\kappa_b ,\kappa_a ) \ .
\end{eqnarray}
In the case $J=0$ there is only one nonvanishing coefficient
$S_{01}(\kappa_a,\kappa_b)=C_0(-\kappa_b,\kappa_a)$. 

It can be immediate seen that $C_J(\kappa_b ,\kappa_a ) \propto \Delta(j_a,j_b,J)$, where $\Delta$
denotes the triangular condition. For the coefficients $S_{JL}$ we have
$S_{JL}(\kappa_a,\kappa_b) \propto \Delta(j_a, j_b, J)$ and
$S_{JL}(\kappa_a,\kappa_b) \propto \Pi(l_a, l_b, L)$.

We note several useful symmetry relations of the angular coefficients:
\begin{equation}\label{s2}
  C_J(\kappa_a ,\kappa_b) = C_J(-\kappa_a ,-\kappa_b) = (-1)^{j_a-j_b} C_J(\kappa_b,\kappa_a )\,,
\end{equation}
\begin{equation}\label{s1}
  S_{JL}(\kappa_a,\kappa_b) = (-1)^{J+L+1} (-1)^{j_b-j_a} S_{JL} (\kappa_b,\kappa_a)\,.
\end{equation}

Several specific values of the angular coefficients relevant for this study are
\begin{eqnarray} 
S_{10}(-1,-1) = \sqrt{2}\,, \ \ S_{10}(1,1) = -\frac{\sqrt{2}}{3}\,,\nonumber \\
S_{10}(-1,2) = 0\,, \ \ S_{10}(1,-2) = \frac43 \,.
\end{eqnarray}

\bibliography{refs}

\end{document}